\documentclass[reprint,floatfix,amsmath,amssymb,aps,prl,nolongbibliography]{revtex4-2}

\usepackage[english]{babel}

\usepackage[hidelinks=true]{hyperref}
\hypersetup{
    colorlinks,
    linkcolor={blue!80!black},
    citecolor={blue!80!black},
    urlcolor={blue!80!black}
}

\usepackage{setspace}
\usepackage{caption}
\usepackage{siunitx}
\usepackage{subcaption}
\usepackage{graphicx}
\usepackage{dcolumn}
\usepackage{bm}
\usepackage{gensymb}
\usepackage{lipsum}

\usepackage{xcolor}
\pagecolor{white}

\begin{document}

\title{Nonmonotonic diffusion in sheared active suspensions of squirmers}

\author{Zhouyang Ge$^{1,2}$}\altaffiliation{Contact author: zhoge@nju.edu.cn. Present address: School of Materials Science and Intelligent Engineering, Nanjing University, Suzhou 215163, China.}
\author{John F.~Brady$^3$}
\author{Gwynn J.~Elfring$^1$}\altaffiliation{Contact author: gelfring@mech.ubc.ca}

\affiliation{$^1$Department of Mechanical Engineering and Institute of Applied Mathematics, University of British Columbia, Vancouver V6T 1Z4, BC, Canada}
\affiliation{$^2$FLOW, Department of Engineering Mechanics, KTH Royal Institute of Technology, 100 44 Stockholm, Sweden}
\affiliation{$^3$Division of Chemistry and Chemical Engineering, California Institute of Technology, Pasadena, California 91125, USA}


\begin{abstract} 
We investigate how shear influences the dynamics of active particles in dilute to concentrated suspensions.
Using apolar active suspensions of squirmers as model systems, we show how their long-time diffusive dynamics can surprisingly slow down and vary nonmonotonically with the shear rate arising from an interplay between the activity-induced persistent motion and shear-induced reorientation and diffusion.
Further simulations of self-propelled particles with tunable persistence exhibit richer dynamics and confirm the observed coupling, suggesting that nonmonotonic diffusion may be a general feature of fluids endowed with an underlying microstructure and large persistence.
Our results reveal a nonlinear effect of shear on diffusion in active suspensions, elucidate how internal and external forcing interact, and provide new possibilities to modulate transport in active fluids.
\end{abstract}

\maketitle

Suspending particles able to inject energy in a fluid provides a means to drive the fluid out of equilibrium internally. 
Such \emph{active suspensions}, or active fluids, have been the subject of intense study in recent decades, showing rich phenomena unseen in their passive, equilibrium counterparts \citep{marchetti_review2013, gompper2020}.
As a typical example, bacterial suspensions have been found to display enhanced diffusion \citep{WL2000, Kim2004, Chen2007, Mino2011, Patteson2016, Peng2016} and turbulent-like chaotic motion \citep{Dombrowski2004, Sokolov2007, Wensink2012, Dunkel2013, Peng2021bacterial} in the absence of any external flow, as well as upstream swimming \citep{Hill2007upstream, Kaya_Koser2009, Kaya_Koser2012} and superfluid-like rheology \citep{lopez2015turning, Saintillan2018, Chui2021rheology} under pressure or shear-driven flows.
Understanding how various internal and external driving mechanisms interact across spatial and temporal scales is important for predicting and controlling the dynamics and mechanics of active fluids in complex conditions, yet a detailed knowledge of the interplay has not been established.

One of the simplest ways to drive a fluid externally is to \emph{shear} it at a steady, homogeneous shear rate, $\dot\gamma$.
When sheared, passive noncolloidal suspensions undergo \emph{shear-induced diffusion} (SID) in the plane perpendicular to the mean flow \citep{Eckstein1977, leighton_acrivos_1987, breedveld1998measurement, sierou2004shear}.
SID arises from fluid-mediated interparticle collisions, and dimensional analysis suggests that the resulting diffusion coefficient, $D_s$, scales as $\dot\gamma a^2 f(\phi)$, where $a$ is the particle radius and $f(\phi)$ a function of the particle volume fraction, $\phi$ \citep{Acrivos1995bingham}.
The precise form of $f(\phi)$ depends on the physical regime as well as the shape and surface properties of the particles \citep{da1996shear, drazer_koplik_khusid_acrivos_2002, Zhang2023shear, Zhang2024shear, Athani2024shear}.
However, regardless of the details, one expects $D_s$ to increase with $\dot\gamma$ because particles collide more frequently at higher $\dot\gamma$.

In this Letter, we investigate the interplay between shear and activity, and examine whether increasing shear rate enhances diffusion in active fluids.
To this end, we use a recently modified Fast Stokesian Dynamics (FSD) method \citep{fiore2019fast, Ge_Elfring_2025} to perform large-scale hydrodynamic simulations of dilute to concentrated active suspensions ($\phi \in [10\%,40\%]$). 
Specifically, we first consider a simple model -- \emph{apolar} active suspensions of ``shakers," particles that are active but not self-propelling, which collectively display \emph{activity-induced hydrodynamic diffusion} \citep{Ge_Elfring_2025}.
The dynamics can be characterized by a diffusion coefficient, $D_0$, allowing for a P\'eclet number, Pe $\equiv \dot\gamma a^2/D_0$, to be defined to measure the relative effect of shear versus activity.
Extensive simulations of thousands of shakers at different Pe and $\phi$ show that such suspensions always reach diffusive dynamics at long times; however, surprisingly, the effective diffusion coefficient, $D$, can \emph{drop below} $D_0$ for a range of Pe when $\phi$ is small. 
For example, $D/D_0 \approx 0.4$ when Pe $\approx 100$ and $\phi=10\%$.
As Pe increases further, $D/D_0$ rises again, with shear eventually dominating, resulting in \emph{nonmonotonic} diffusion.
We analyze this behavior in detail and trace its physical origin to an activity-induced persistent motion and a shear-induced negative velocity autocorrelation, both of which are more pronounced at lower volume fractions. 
To confirm and generalize our understanding, we then consider self-propelled particles with \emph{tunable} persistence, which display nonmonotonic diffusion even in \emph{concentrated} suspensions, provided that the persistence is large enough.
Overall, our results demonstrate a nonlinear effect of shear on diffusion in active suspensions, elucidate how interactions between internal and external forcing influence these complex fluids, and suggest new ways to control mixing and mass transport in active fluids.

\begin{figure*}[t]
  \centering
  \includegraphics[width=\textwidth]{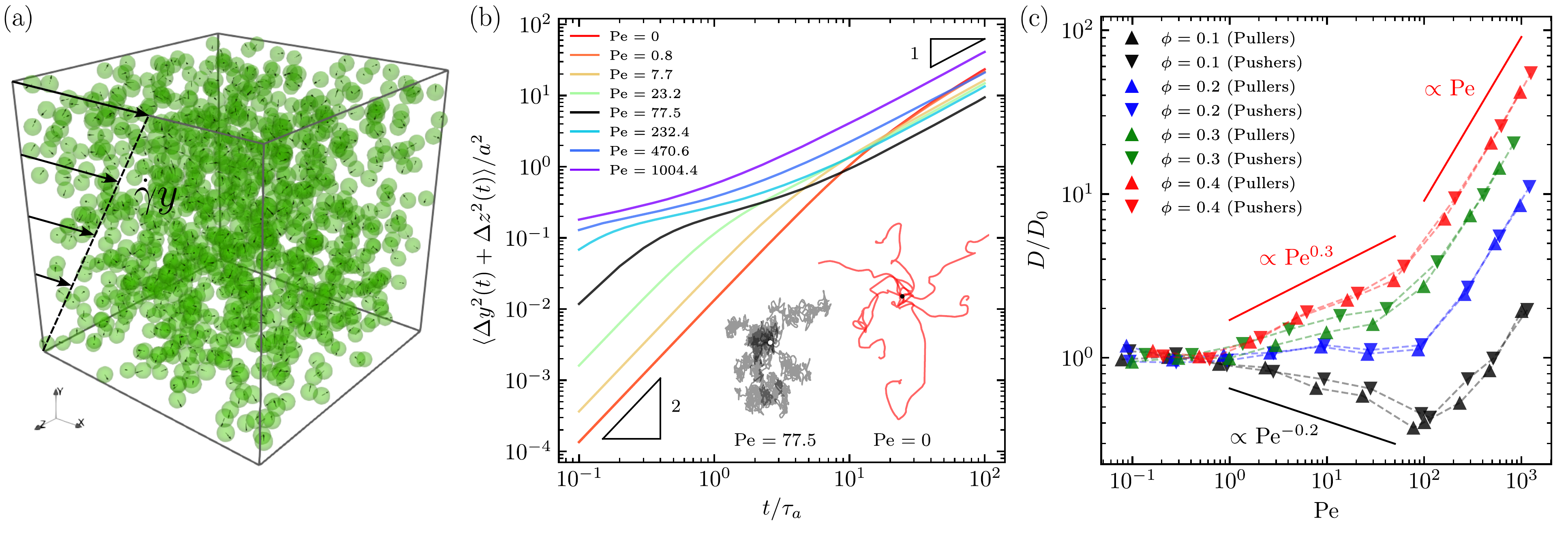}
  \caption{Nonmonotonic diffusion in sheared active suspensions.
  (a) Illustration of 1024 particles at 10\% volume fraction ($\phi$) under shear.
  (b) Mean square displacements in the $yz$-plane for pulling shakers ($\phi=10\%$) at different P\'eclet numbers, Pe. The inset shows a few sample trajectories at Pe $=0$ and 77.5.
  (c) Relative diffusion coefficient, $D/D_0$, against Pe for pulling or pushing shakers at different $\phi$. Slopes are theoretical estimates (see text).}
  \label{fig:D}
\end{figure*}

\emph{Models and methods}---We model active particles as ``squirmers," spheres with a prescribed axisymmetric slip velocity \citep{lighthill1952, blake1971, Pedley2016}.
The surface slip at position $\bm r$ from the sphere's center may be expressed as
\begin{align}
    \bm u_\text{s} = -\bm U_\text{s} - \bm \Omega_\text{s} \times \bm r + 
    \bm r \cdot \bm E_\text{s} + \dots,
\end{align}
where $\bm U_\text{s}$ is the self-propulsion velocity,
$\bm \Omega_\text{s}$ the self spin, 
$\bm E_\text{s}$ the self strain rate,
and dots denote terms involving higher order irreducible tensors of $\bm r$ \citep{elfring2022active}.
Specifically,
\begin{align}
\bm U_\text{s} = \frac{2}{3}B_1 \bm p, \  
\bm \Omega_\text{s} = \frac{C_1}{a^3} \bm p, \  
\bm E_\text{s} = -\frac{3}{5} \frac{B_2}{a} \big(\bm p \bm p-\frac{1}{3} \bm I \big),
\end{align}
where $B_1$ and $B_2$ are the leading-order squirming modes, 
$C_1$ a coefficient associated with azimuthal slip, 
$\bm p$ the particle director,
and $\bm I$ an identity tensor \citep{elfring2022active}.
Setting $B_1=0$ yields an immotile squirmer known as a \emph{shaker}.

Although shakers are individually immotile, they swim collectively due to \emph{hydrodynamic interactions} (HIs). 
To leading order, the HIs arise from $\bm E_\text{s}$ if $B_2 \ne 0$: $B_2>0$ corresponds to \emph{pullers}, while $B_2<0$ \emph{pushers}. 
Therefore, the ratio $\beta \equiv B_2/B_1$ indicates the type of swimmer and the relative importance of HIs and self-propulsion. 
In this work, we first take $\beta \to \pm \infty$ to amplify the effect of HIs, then consider neutral squirmers with $\beta=0$ and additional rotational diffusion of $\bm p$ on each particle, effectively modeling \emph{active Brownian particles} (ABPs). 
Higher-order squirming modes and the azimuthal slip ($C_1$) are neglected, as is common in the literature \citep{ishikawa2006hydrodynamic, Ishikawa2010diffusion, Delmotte2018}.

The dynamics of a suspension of squirmers is described by the active Stokesian Dynamics framework \citep{elfring2022active}.
For athermal particles, the governing equations can be written compactly as
\begin{align} \label{eq:asd}
    \mathbf U = \mathbf U_\infty + \mathbf U_\text{s} + \mathcal{R}_\mathrm{FU}^{-1} \bm\cdot
    \big( \mathcal R_\text{FE} \bm:(\mathbf E_\infty - \mathbf E_\text{s}) + \mathbf F_\text{ext} \big),
\end{align}
where the velocities ($\mathbf U, \mathbf U_\infty, \mathbf U_\text{s}$), strain rates ($\mathbf E_\infty, \mathbf E_\text{s}$) and force ($\mathbf F_\text{ext}$) are all $N$-tuples, with elements defined at the particle centers, e.g., ${\mathbf U} \equiv (\bm U_1 \ \bm U_2 \ ... \ \bm U_N)^T$, and ${\mathcal R}_\text{FU}$ and ${\mathcal R}_\text{FE}$ are the resistance tensors coupling the force and velocity moments of all $N$ particles.
Note how individual particle activities couple to the large-scale flows through HIs. 
As illustrated in Fig.~\ref{fig:D}(a), we apply a shear in the $x$ direction, $\bm U_\infty (x,y,z)=(\dot\gamma y,0,0)$, and impose a short-range repulsive force $\bm F_\text{ext}$ between the particles to model excluded-volume interactions; see the Supplemental Material (SM) \footnote{See Supplemental Material at [url] for a detailed description of our simulation method, setup and verification, which also includes Refs.~\citep{Ge2022, Israelachvili_book, Lees_Edwards_1972, Mari_2014JOR, Beenakker1984, ladd1990hydrodynamic, Lastra2013subdiffusive}.} for further details.
Eq.~(\ref{eq:asd}) is then solved using FSD \citep{fiore2019fast} for $N=1024$ at different Pe and $\phi$.

\emph{Shakers}---%
Recently, we showed that apolar active suspensions of shakers, pulling or pushing, could undergo activity-induced diffusion without any external flow \citep{Ge_Elfring_2025}.
The effective diffusion coefficient was found to be $D_0=(a^2/\tau_a)g(\phi)$, where $\tau_a \equiv a/|B_2|$ is the activity time scale, and $g(\phi) \sim \phi^{1/5}$ for $\phi \lessapprox 10\%$ and reduces sharply with $\phi$ beyond.
Furthermore, the short-time dynamics was ballistic and could last over tens of $\tau_a$; see the Pe $=0$ case in Fig.~\ref{fig:D}(b).
This \emph{persistent} motion, therein due to HIs between the active particles, is the main reason that active fluids display an enhanced diffusion \citep{WL2000, Mino2011, Ge_Elfring_2025}.

\begin{figure*}[t]
  \centering
  \includegraphics[width=\textwidth]{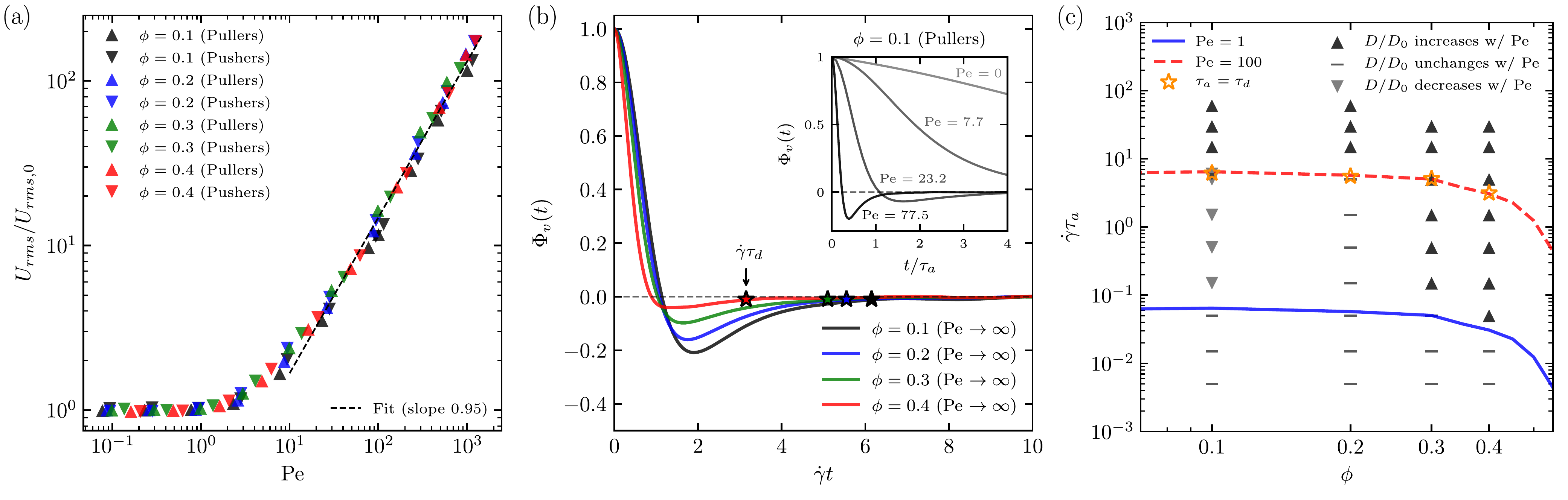}
  \caption{Short-time dynamics and phase diagram. 
  (a) Root-mean-square particle speeds under shear, normalized by their values without shear at all Pe and $\phi$.
  (b) Normalized velocity autocorrelation for passive particles at different $\phi$ (markers denote $\Phi_v(\dot\gamma \tau_d) = -10^{-2}$) and for pulling shakers under different Pe at $\phi=10\%$ (inset).
  (c) Phase diagram of all simulation results for shakers, divided by Pe $=1$ and 100 into three diffusive regimes with different scalings.}
  \label{fig:rms}
\end{figure*}

Under shear, the particle dynamics is anisotropic and not diffusive in the flow direction ($x$); thus, we analyze the mean square displacement (MSD) in the $yz$-plane, $\langle \Delta y^2(t) + \Delta z^2(t) \rangle$, where $\langle \cdot \rangle$ averages over all particles and time.
Fig.~\ref{fig:D}(b) shows the MSDs for pulling shakers under different Pe at $\phi=10\%$ (see SM for results of pushers and at other $\phi$ \citep{Note1}).
We observe that the short-time MSD increases with Pe, similar to the behavior of passive suspensions \cite{sierou2004shear}; however, surprisingly, the long-time dynamics can be \emph{suppressed} by shear, with sustained \emph{subdiffusive} motion, resulting in smaller MSDs at larger Pe.
The reduced diffusion is also evident from direct visualization of individual particle trajectories. 
As shown in the inset in Fig.~\ref{fig:D}(b), increasing Pe could lead to more compact and less smooth trajectories at long times.

We extract the diffusion coefficient, $D$, by fitting the long-time MSD according to $\langle \Delta y^2(t) + \Delta z^2(t) \rangle = 4D(t-t_0)$, where $t_0$ is generally nonzero due to nondiffusive dynamics at short times.
Fig.~\ref{fig:D}(c) shows the relative diffusivity, $D/D_0$, across four decades of Pe. 
Clearly, $D/D_0$ is nonmonotonic at $\phi=10\%$, with a minimum of about 0.4 reached at Pe $\approx 100$, consistent with the observation in Fig.~\ref{fig:D}(b).
At higher concentrations, the nonmonotonic behavior gradually disappears, and the scaling between $D/D_0$ and Pe depends on $\phi$ (see analysis later).
However, in all cases, we observe $D/D_0 \approx 1$ for Pe $\lessapprox 1$ and $D/D_0 \propto$ Pe for Pe $\gtrapprox 100$. 
The same scalings apply to both pullers and pushers, as is the $\phi$-dependence of $D_0$ without shear (c.f.~Fig.~\ref{fig:D-phi}) \citep{Ge_Elfring_2025}, implying similar but still irreversible macroscopic dynamics under time reversal.

To explain the nonmonotonic diffusion, we examine the short-time dynamics characterized by the \emph{instantaneous} particle velocities, projected onto the $yz$-plane as in the MSD calculation.
Fig.~\ref{fig:rms}(a) shows the root-mean-square particle speeds under shear, $U_{rms}$, normalized by their values without shear, $U_{rms,0}$, for pullers and pushers at different $\phi$.
Remarkably, all data collapse onto a single curve, showing universal scalings with $U_{rms}/U_{rms,0} \approx 1$ for Pe $\lessapprox 1$ and $U_{rms}/U_{rms,0} \propto$ Pe for Pe $\gtrapprox 10$.
The absence of any $\phi$-dependence suggests that the relative change of the short-time dynamics is controlled by shear alone, despite the underlying activity-induced dynamics, $U_{rms,0}$ and $D_0$, depending nonmonotonically on $\phi$ \citep{Ge_Elfring_2025}.

The instantaneous velocities decorrelate over time and can be quantified by a velocity autocorrelation function, $\Phi_v(t) \equiv \langle \bm U(\tau) \cdot \bm U(\tau+t) \rangle/U_{rms}^2$, where $\langle \cdot \rangle$ averages over all particles and time.
Fig.~\ref{fig:rms}(b) shows $\Phi_v(t)$ for pulling shakers under different Pe at $\phi=10\%$ and passive particles (Pe $\to \infty$ as activity is turned off) at different $\phi$.
We observe faster decorrelation as Pe or $\phi$ increases and even a negative $\Phi_v(t)$ at sufficiently high Pe.
Such \emph{anticorrelation} has been noted in earlier simulations \citep{drazer_koplik_khusid_acrivos_2002, sierou2004shear} and results from either hydrodynamic or excluded-volume interactions.
Under shear, pairwise HIs render transient particle displacements that necessarily become zero at equal time after the encounter \citep{batchelor1972hydrodynamic}; thus, the velocities change sign and $\Phi_v(t) <0$. 
This effect is weakened at higher $\phi$ due to many-body HIs and more frequent collisions; however, particles tend to oscillate as they collide (see supplemental videos), again making $\Phi_v(t) <0$. 
The time scale of these shear-induced interactions is proportional to $\dot\gamma^{-1}$ and reduces with $\phi$, c.f.~Fig.~\ref{fig:rms}(b).
The negative velocity autocorrelation is the reason for the subdiffusive motion observed in Fig.~\ref{fig:D}(b).

We now explain the observed diffusion.
Fig.~\ref{fig:rms}(c) shows the phase diagram of all shaker simulations controlled by two parameters, $\phi$ and $\dot\gamma \tau_a = g(\phi)$Pe. 
As discussed, there are three diffusive regimes separated by Pe $\approx 1$ and 100, with different scalings between $D/D_0$ and Pe in each.
Specifically, Pe $<1$ corresponds to $a^2/D_0 < \dot\gamma^{-1}$, i.e., the activity-induced diffusion occurs faster than the shear-induced, thus we observe the former, $D=D_0$.
When Pe $>1$, both activity and shear are at play, resulting in $\phi$-dependent scalings, including a reduced diffusion at low $\phi$ due to the shear-induced velocity anticorrelation. 
However, at sufficiently large Pe, the dynamics must be controlled by shear, $D =D_s \propto \dot\gamma a^2$, leading to $D/D_0 \propto$ Pe.
Here, the threshold Pe corresponds to $\tau_d = \tau_a$, where $\dot\gamma\tau_d$ is a $\phi$-dependent velocity decorrelation time, i.e., the time $\Phi_v(t)$ \emph{returns} to zero from below, c.f.~Fig.~\ref{fig:rms}(b).
When $\tau_d < \tau_a$, diffusion occurs due to shear because even the shortest activity time scale ($\tau_a$) is longer than the diffusive time of shear ($\tau_d$).
Indeed, Fig.~\ref{fig:rms}(c) shows that $\tau_a = \tau_d$ approximates Pe $=100$. 
Furthermore, we can also estimate the scaling between $D/D_0$ and Pe in the intermediate Pe range.
Suppose $D/D_0 \sim$ Pe$^\nu$, and let $D=D_0$ (or $D_s$) at Pe $=$ Pe$_0$ (or Pe$_s$); then, $\nu=\log(D_s/D_0)/\log(\textrm{Pe}_s/\textrm{Pe}_0)$.
Using Pe$_0=1$, Pe$_s=100$, and $D_s/D_0 = 0.4$ (or 4.0), we obtain $\nu=-0.2$ (or 0.3)  at $\phi=0.1$ (or 0.4).
Fig.~\ref{fig:D}(c) verifies our estimates.

\emph{ABPs}---One may wonder whether the nonmonotonic diffusion observed here is a general feature of fluids with persistent internal motion.
After all, if it is observable in apolar active suspensions, the effect may be stronger for \emph{polar} particles with a larger persistence.
To test this hypothesis, we perform additional simulations of ABPs, a widely studied model \citep{Romanczuk2012ABPs} for \emph{self-propelled} particles, e.g., self-phoretic colloids \citep{Golestanian2009}.
Specifically, we simulate neutral squirmers with $B_1 > 0$ and $B_2=0$, and impose a uniform rotational diffusion, $d_r$, to effect angular noise in the particle dynamics, $\langle \bm \Omega_\text{s}(t) \cdot \bm \Omega_\text{s}(t^\prime) \rangle = 2d_r \delta(t-t^\prime)$, where $\delta(t)$ is the Dirac delta.
Translational noise is neglected as done typically \citep{Fily2012}.
Without shear, our ABPs self-propel with a constant speed, $U_s = \frac{2}{3}B_1$, unless they collide. 
Over time, collision and orientation relaxation render the motion diffusive, with $D_{0,\textrm{ABPs}} \to \frac{1}{3}U_s^2 \tau_r$ as $\phi \to 0$, where $\tau_r \equiv \frac{1}{2} d_r^{-1}$.
The persistence length, $\ell_p \equiv 3D_{0,\textrm{ABPs}}/U_{rms}$, is \emph{tunable} and capped at $\ell_{p,0} \equiv U_s\tau_r$ as $\phi \to 0$ \citep{ Note1}.

\begin{figure}[t]
  \centering
  \includegraphics[width=0.95\columnwidth]{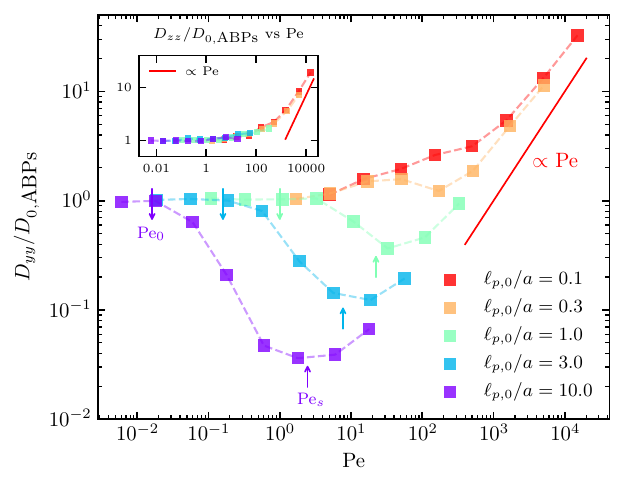}
  \caption{Diffusivity of ABPs with different persistence length ($\ell_p$) at $\phi=10\%$.
  The main plot shows the relative diffusivity in the $y$ direction; the inset in $z$.
  The downarrows denote Pe$_0 \equiv \min(1,(a/\ell_p)^2)$; the uparrows Pe$_s \equiv \dot\gamma_s a^2/D_{0,\textrm{ABPs}}$, where $\dot\gamma_s \equiv \dot\gamma \tau_d/\min(\tau_s, \tau_r)$.}
  \label{fig:ABPs}
\end{figure}

Fig.~\ref{fig:ABPs} shows the separate diffusivities in $y$ and $z$ directions, $D_{yy}$ and $D_{zz}$, relative to $D_{0,\textrm{ABPs}}$, for different $\ell_p \approx \ell_{p,0}$ \citep{ Note1} at $\phi=10\%$ under shear.
The diffusivities are extracted in the same way as for shakers.
For $D_{yy}$, we observe the expected effect of $\ell_p$, i.e., a larger persistence leads to a larger drop in diffusivity, which can be more than 90\% if $\ell_p/a > 3$ (as a reference, $\ell_p/a \sim O(1)$ for shakers at $\phi=10\%$ \citep{Ge_Elfring_2025}).
Conversely, the nonmonotonic effect disappears when $\ell_p/a \ll 1$, and we have verified that the diffusivity of passive colloids ($\ell_p=0$) strictly increases with shear \citep{Note1}.
The Pe at which motility ceases (or shear begins) to dominate, Pe$_0$ (or Pe$_s$), can be estimated similarly as in the case of shakers:
motility dominates when $\max(a^2/D_{0,\textrm{ABPs}}, \ell_p^2/D_{0,\textrm{ABPs}}) < \dot\gamma^{-1}$, or Pe $<\min(1,(a/\ell_p)^2) \equiv$ Pe$_0$,
while shear dominates when $\tau_d < \min(\tau_s, \tau_r)$, or Pe $> \dot\gamma_s a^2/D_{0,\textrm{ABPs}} \equiv$ Pe$_s$, where $\tau_s \equiv a/U_s$ and $\dot\gamma_s \equiv \dot\gamma \tau_d/\min(\tau_s, \tau_r)$.
Here, a major difference between ABPs and shakers is that $D_{zz}/D_{0,\textrm{ABPs}}$ always increases with Pe, and the data at different $\ell_p$ collapse, scaling as SID. 
The reason for the contrasting behaviors is that ABPs hardly interact.
As shear rotates all particles around the $z$ axis with a constant speed $\Omega_\infty = \dot\gamma/2$, ABPs oscillate in the $xy$-plane but remain persistent along $z$ (manifest in the MSDs \citep{Note1}).
Consequently, $D_{zz}/D_{0,\textrm{ABPs}}$ can decouple from $D_{yy}/D_{0,\textrm{ABPs}}$ over a large range of Pe, leading to highly anisotropic diffusion.


\begin{figure}[t]
  \centering
  \includegraphics[width=0.85\columnwidth]{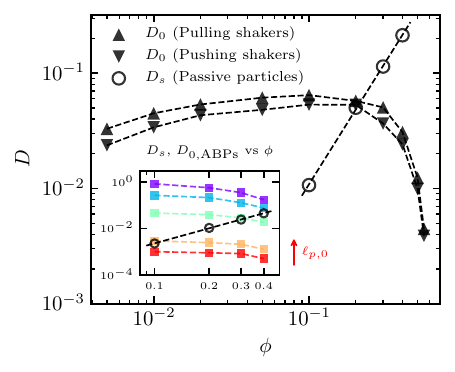}
  \caption{Comparison of the different intrinsic and shear-induced diffusivities (in the same unit, $a^2/\tau_a$). 
  The main plot compares the $D_0$ for pulling or pushing shakers with $D_s$ for passive particles at $\dot\gamma_s (\phi=0.2)$. 
  The inset compares $D_{0,\textrm{ABPs}}$ at different $\ell_{p,0}$ (same color legend as in Fig.~\ref{fig:ABPs}) with $D_s$ at $\dot\gamma_s (\phi=0.4, \ell_{p,0}/a=10)$.}
  \label{fig:D-phi}
\end{figure}


\emph{Discussions}---%
The results of our simulations depict a general physical picture of the interplay between shear and activity.
When shearing a fluid with an underlying persistence, arising from either activity-induced HIs (as in the case of shakers), self-propulsion (ABPs), or a combination of the two (squirmers in general), the rotational effect of shear reduces the persistent motion in the shear plane. 
For a single ABP without any SID (as $\phi \to 0$), this leads to a \emph{monotonically} reducing $D_{yy}/D_{0,\textrm{ABPs}}$ with $\dot\gamma \tau_r$ \citep{Sandoval_2014, Takatori_Brady2017}.
At finite $\phi$, both shear-induced rotation and collision are at play, and the dynamics depends on the competition of the intrinsic diffusion and SID.
If $D_s(\dot\gamma_s,\phi) < D_0(\tau_a, \phi)$, where $\dot\gamma_s(\phi)$ is the shear rate above which SID dominates, we would expect a reduced diffusion as Pe increases from Pe$_0$ to Pe$_s$. 
Conversely, if $D_s(\dot\gamma_s,\phi) > D_0(\tau_a, \phi)$, an enhanced diffusion is expected. 
Fig.~\ref{fig:D-phi} shows that $D_0$ of both pulling and pushing shakers cross $D_s$ at $\phi \approx 20\%$, which is why $D/D_0$ is nearly independent of Pe up to Pe$_s$ at $\phi = 20\%$; c.f.~Fig.~\ref{fig:D}(c).
Furthermore, the inset in Fig.~\ref{fig:D-phi} suggests that diffusion could be suppressed even in \emph{concentrated} suspensions, since $D_s(\dot\gamma_s,\phi) < D_{0,\textrm{ABPs}}(\tau_s,\tau_r,\phi)$ at sufficiently large $\ell_p$, 
which we indeed observe at $\phi=40\%$ and $\ell_{p,0}/a=10$ \citep{Note1}.
Therefore, we expect nonmonotonic diffusion to be a general feature when shearing fluids endowed with an underlying microstructure and large internal persistence.
Finally, there is currently a growing interest in the relation of the dynamics, structure, and rheology of active fluids, which has great practical implications but is still not well understood \citep{Fielding2011, Takatori_Brady2017, Guo_PNAS_2018, Martinez2020, John_review_2025, martin2025mips}.
Our study may shed some light on this problem.

\emph{Acknowledgments}---%
We are grateful to Boyuan Chen for help in the code development, and thank Eva Kanso, Tingtao Zhou, and the anonymous referees for useful comments.
Z.G.~acknowledges support from the Swedish Research Council under Grant No.~2021-06669VR.

\emph{Data availability}---%
The source code to generate all data reported in this article is publically available \footnote{\url{https://github.com/GeZhouyang/FSD}}.


\bibliographystyle{apsrev4-2}
\bibliography{main}

\end{document}


\title{Supplemental Material for ``Nonmonotonic diffusion in sheared active suspensions of squirmers"}

\author{Zhouyang Ge$^{1,2}$}\altaffiliation{Contact author: zhoge@nju.edu.cn. Present address: School of Materials Science and Intelligent Engineering, Nanjing University, Suzhou 215163, China.} 
\author{John F.~Brady$^3$}
\author{Gwynn J.~Elfring$^1$}

\affiliation{$^1$Department of Mechanical Engineering and Institute of Applied Mathematics, University of British Columbia, Vancouver V6T 1Z4, BC, Canada}
\affiliation{$^2$FLOW, Department of Engineering Mechanics, KTH Royal Institute of Technology, 100 44 Stockholm, Sweden}
\affiliation{$^3$Division of Chemistry and Chemical Engineering, California Institute of Technology, Pasadena, California 91125, USA}


\maketitle

\tableofcontents
	
\newpage

\section{Models and methods}

\subsection{Fast Stokesian Dynamics}
\label{sec:FSD}

The results in the main article were obtained from large-scale hydrodynamic simulations using the Fast Stokesian Dynamics (FSD) method \citep{fiore2019fast}, modified to include particle activities according to the active Stokesian Dynamics framework \citep{elfring2022active}.
Our modification was reported in detail in a previous work \cite{Ge_Elfring_2025}, which did not include Brownian motion.
Below, we provide a brief summary of the complete formulation.

Assuming small, inertialess particles suspended in a viscous fluid at low Reynolds numbers, the external force moments on any particle must be balanced by their hydrodynamic counterparts, which can be linearly related to the velocity moments on \emph{all} particles through hydrodynamic resistance tensors.
In the presence of external flow and Brownian motion, this leads to 
\begin{equation} \label{eq:active-sd}
    \mathbf U =  \mathbf U_\infty + \mathbf U_\text{s} + \mathbf U_\text{B} + \mathcal{R}_\mathrm{FU}^{-1} \bm\cdot
    \big(\mathcal R_\text{FE} \bm: (\mathbf E_\infty - \mathbf E_\text{s}) + \mathbf F_\text{ext} \big),
\end{equation}
where the velocities ($\mathbf U, \mathbf U_\infty, \mathbf U_\text{s}, \mathbf U_\text{B}$), strain rates ($\mathbf E_\infty, \mathbf E_\text{s}$) and force ($\mathbf F_\text{ext}$) are all $N$-tuples, with elements defined at the particle centers, e.g., ${\mathbf U} \equiv (\bm U_1 \ \bm U_2 \ ... \ \bm U_N)^T$, and ${\mathcal R}_\text{FU}$ and ${\mathcal R}_\text{FE}$ are the resistance tensors coupling the force and velocity moments of all $N$ particles.
We apply an external shear flow with shear rate $\dot\gamma$ along the $x$ direction. 
The ambient velocity $\bm U_\infty$ at position $(x,y,z)$ contains a straining motion $\bm E_\infty$ and a rotation $\bm \Omega_\infty$, given as 
\begin{equation}
  \bm U_\infty =
  \begin{bmatrix}
    \dot\gamma y \\ 0 \\ 0
  \end{bmatrix}, \quad
  \bm E_\infty =
  \begin{bmatrix}
  0 & \dot\gamma/2 & \ 0 \\
  \dot\gamma/2 & 0 & \ 0 \\
  0 & 0 & \ 0
\end{bmatrix}, \quad
  \bm \Omega_\infty =
  \begin{bmatrix}
    0 \\ 0 \\ -\dot\gamma/2
  \end{bmatrix}.
\end{equation} 
Note that the velocity defined on each particle in Eq.~\eqref{eq:active-sd} includes both the linear and angular components.
The same applies to the force on each particle.

For a squirmer, the self-propulsion velocity $\bm U_\text{s}$ and active strain rate $\bm E_\text{s}$ are proportional to its squirming modes $B_1$ and $B_2$, respectively,
\begin{align}
    \bm U_\text{s} = \frac{2}{3}B_1 \bm p, \quad  
    \bm E_\text{s} = -\frac{3}{5} \frac{B_2}{a} (\bm p \bm p-\frac{1}{3} \bm I),
\end{align}
where $a$ is the particle radius, $\bm p$ its orientation vector, and $\bm I$ an identity tensor.
Both modes could induce hydrodynamic interactions between squirmers; however, we only include the leading-order term due to $\bm E_\text{s}$, as the contribution from $\bm U_\text{s}$ appears in the next order \citep{elfring2022active}.
Azimuthal slips, e.g., $\bm \Omega_\text{s}$, can also be included as mentioned in the main article.

To model excluded-volume interactions, we impose a short-range repulsive force between a pair particles separated by a distance vector $\bm r_{ij}$,
\begin{align} \label{eq:F_ext}
    \bm F_\text{ext}^{(ij)} = - F_0 \exp (-\kappa h_{ij}) \frac{\bm r_{ij}}{|\bm r_{ij}|},
\end{align}
where $F_0$ is the maximal repulsive force, $\kappa$ an inverse characteristic length, and $h_{ij}=|\bm r_{ij}|-2a$ the surface gap.
The magnitude of the force decays rapidly with distance and is truncated at a few particle radii. Physically, $\bm F_\text{ext}$ resembles the electrostatic repulsion between colloids suspended in liquids \citep{Israelachvili_book, Mari_2014JOR}.
Numerically, it also serves as a convenient means to prevent particle overlaps that may occur when integrating the particle motion in a flow \citep{Mari_2014JOR, Ge_Elfring_2025, Ge2022}.

The third term on the right-hand side of Eq.~\eqref{eq:active-sd}, which is neglected in Eq.~(3) in the main article (since our active particles are athermal), corresponds to Brownian motion.
Specifically, 
\begin{equation} 
    \mathbf U_\text{B} = \sqrt{2 k_\text{B} T} \mathcal {R}_\text{FU}^{-\frac{1}{2}} \bm\cdot \odv{\mathbf W}{t} + 
    k_\text{B} T \bm\nabla \bm\cdot \mathcal {R}_\text{FU}^{-1},
\end{equation}
where $k_\text{B}$ is the Boltzmann's constant, $T$ the absolute temperature, and $\mathbf W(t)$ the Wiener process (a $N$-tuple similar to $\mathbf U$).
Because $\mathbf U_\text{B}$ is related to the square root and divergence of the inverse resistance tensor, sampling Brownian motion in the presence of hydrodynamic interactions (HIs) is expensive. 
This motivated the development of fast simulation methods, such as the FSD \citep{fiore2019fast}.

We use FSD to solve a modified form of Eq.~\eqref{eq:active-sd} formulated as follows.
As in the conventional Stokesian Dynamics \citep{sd1988}, FSD decomposes the hydrodynamic resistance tensor into a far-field and a near-field part, where the former is obtained by inverting a truncated multipole expansion of the Stokes flow induced by all particles,  and the latter is considered by including the pairwise lubrication minus the duplicated parts in the inverse of the mobility.
Specifically, this can be expressed as
\begin{equation} \label{eq:decomp}
  \mathcal R_\text{FU}= \mathcal{B}^T \bm\cdot (\mathcal{M}^\text{ff})^{-1} \bm\cdot \mathcal{B} + \mathcal R_\text{FU}^\text{nf}, \quad
  \mathcal R_\text{FE}= \mathcal{B}^T \bm\cdot (\mathcal{M}^\text{ff})^{-1} \bm\cdot \mathcal{C} + \mathcal R_\text{FE}^\text{nf},
\end{equation}
where $\mathcal R^\text{nf}_\text{FU}$ and $\mathcal R^\text{nf}_\text{FE}$ are the near-field resistance tensors, relating velocity moments to the near-field hydrodynamic force moments (containing force and torque),
\begin{align} \label{eq:lub}
    \mathbf F^\text{nf} = - \mathcal R_\text{FU}^\text{nf} \bm\cdot (\mathbf U -\mathbf U_\infty - \mathbf U_\text{s} - \mathbf U_\text{B}) - 
    						       \mathcal R_\text{FE}^\text{nf} \bm: (\mathbf E_\text{s} - \mathbf E_\infty),
\end{align}
and $\mathcal{M}^\text{ff}$ is the far-field mobility tensor relating far-field force moments, $\mathcal{F}^\text{ff} \equiv (\mathbf F^\text{ff} \ \mathbf S^\text{ff})^T$, to the velocity moments, $\mathcal{U} \equiv (\mathbf U \ \mathbf E)^T$, through mapping tensors $\mathcal{B}$ and $\mathcal{C}$, 
\begin{equation} \label{eq:maps}
  \mathcal{U} = - \mathcal M^\text{ff} \bm\cdot \mathcal F^\text{ff}, \quad
  \mathcal{B} \bm\cdot \mathbf U  + \mathcal{C} \bm\cdot \mathbf E = \mathcal{U}, \quad
  \mathcal{B}^T \bm\cdot \mathcal{F}^\text{ff} = \mathbf F^\text{ff} .
\end{equation}
The negative signs in Eqs.~(\ref{eq:lub}, \ref{eq:maps}) reflect that the hydrodynamic force is a drag force.

Similarly, the Brownian motion is decomposed into a far-field generalized slip $\mathcal U_\text{B}^\text{ff}$ and a near-field force $\mathbf F_\text{B}^\text{nf}$, in addition to a drift velocity $\mathbf U_\text{drift}$.
These terms are given as
\begin{equation} 
    \mathcal U_\text{B}^\text{ff} = \sqrt{2 k_\text{B} T} \big(\mathcal {M}^\text{ff} \big)^{\frac{1}{2}} \bm\cdot \odv{\mathbf W}{t}, \quad 
    \mathbf F_\text{B}^\text{nf} = \sqrt{2 k_\text{B} T} \big(\mathcal {R}_\text{FU}^\text{nf} \big)^{\frac{1}{2}} \bm\cdot \odv{\mathbf W}{t}, \quad 
    \mathbf U_\text{drift} = k_\text{B} T \bm \nabla \cdot \mathcal {R}_\text{FU}^{-1},
\end{equation}
where $\mathbf W(t)$ has different physical meanings in different terms (as implied), and $\mathcal U_\text{B}^\text{ff}$ includes both the velocities and the strain rates due to the far-field Brownian forces and stresses.

Finally, we group all related tuples 
\begin{equation}  
 \begin{split}
  \hat{\mathcal F}^\text{ff} = \mathcal F^\text{ff} + (\mathcal{M}^\text{ff})^{-1} \bm\cdot \mathcal U_\text{B}^\text{ff}, \quad &
  \mathcal U_* = \mathcal U_\text{B}^\text{ff} + \mathcal C \bm\cdot \mathbf (\mathbf E_\infty - \mathbf E_\text{s}), \\
  \mathbf U_* =  \mathbf U - \mathbf U_\infty - \mathbf U_\text{s} - \mathbf U_\text{drift}, \quad &
  \mathbf F_* = - \mathbf F_\text{ext} - \mathcal R_\text{FE}^\text{nf} \bm: (\mathbf E_\infty - \mathbf E_\text{s}) - \mathbf F_\text{B}^\text{nf},
 \end{split}
\end{equation}
to reformulate Eq.~\eqref{eq:active-sd} into a compact saddle-point equation,
\begin{equation} \label{eq:final}
\begin{bmatrix}
  \mathcal{M}^\text{ff} & \mathcal{B} \\
  \mathcal{B}^T & - \mathcal{R}_\text{FU}^\text{nf}
\end{bmatrix}
\bm\cdot
\begin{bmatrix} 
  \hat{\mathcal F}^\text{ff} \\ 
  \mathbf{U}_*
\end{bmatrix}
=
\begin{bmatrix}
  \mathcal U_* \\ 
  \mathbf{F}_*
\end{bmatrix}.
\end{equation}
Here, the first line represents the far-field mobility relationship, while the second line the force balance.

Eq.~\eqref{eq:final} is the governing equation of our modified, active FSD method.
It can be solved efficiently using the numerical procedure developed in the original FSD on a graphical processing unit.
Once $\mathbf U$ is obtained, the particle positions and orientations are integrated forward in time using the standard Euler-Maruyama method, or simply the Euler method in the absence of Brownian motion or rotational diffusion.
Although the original FSD solver was provided in the supplemental material of Ref.~\cite{fiore2019fast}, the original version contained a number of errors.
The source code of our improved implementation is available at \url{https://github.com/GeZhouyang/FSD}; see Refs.~\cite{fiore2019fast, Ge2022, Ge_Elfring_2025} for further descriptions of the hydrodynamic tensors and numerical techniques.

\subsection{Simulation setup}
\label{sec:setup}

\begin{table}[t]
  \centering
  \setlength{\tabcolsep}{14pt} \renewcommand{\arraystretch}{1.2} 
  \begin{tabular}{cccccccc}
    \hline
    $\phi$   &   $B_1$   &   $B_2$    &   $\tau_a$  \\
    \hline
    10\% -- 40\% & 0 & $\pm 0.05$ & 20\\  
    10\% -- 40\% & 0 & $\pm 0.10$ & 10\\  
    10\% -- 40\% & 0 & $\pm 0.20$ & 5\\  
    \hline
  \end{tabular}
  \caption{Parameters of the shaker simulations, where $\tau_a \equiv a/|B_2|$ is the time scale due to activity. Brownian motion is not included in these simulations ($k_B T=0$).}
  \label{tab:shakers}
\end{table}

\begin{table}[t]
  \centering
  \setlength{\tabcolsep}{14pt} \renewcommand{\arraystretch}{1.2} 
  \begin{tabular}{cccccccc}
    \hline
    $\phi$   &   $B_1$   &  $B_2$ & \xx{$U_s$}  & \xx{$\tau_r$}  & \xx{$\ell_{p,0}$} \\
    \hline
    10\% \xx{-- 40\%} & 0.01 & 0  & \xx{1/150}  & \xx{15}  & 0.1    \\  
    10\% \xx{-- 40\%} & 0.01 & 0  & \xx{1/150}  & \xx{45}  & 0.3    \\  
    10\% \xx{-- 40\%} & 0.05 & 0  & \xx{1/30}   & \xx{30}   & 1.0    \\  
    10\% \xx{-- 40\%} & 0.10 & 0  & \xx{1/15}   & \xx{45}   & 3.0    \\  
    10\% \xx{-- 40\%} & 0.10 & 0  & \xx{1/15}   & \xx{150}  & 10.0    \\  
    \hline
  \end{tabular}
  \caption{Parameters of the ABP simulations, where $U_s \equiv \frac{2}{3}B_1$ is the self-propulsion speed, $\tau_r \equiv \frac{1}{2}d_r^{-1}$ a time scale due to rotational diffusion, and $\ell_{p,0} \equiv U_s\tau_r$ the persistence length as $\phi \to 0$. Brownian motion is not included in these simulations ($k_B T=0$).}
  \label{tab:ABPs}
\end{table}

We simulate $N=1024$ active or passive particles in a cubic box undergoing simple shear.
The particles are all of the same size, and their positions and velocities are governed by the Lees-Edwards periodic boundary condition \citep{Lees_Edwards_1972}.
We normalize all lengths by the particle radius, $a$, all times by the inverse of an unit shear rate, $\tau_\text{ref} \equiv 1/\dot\gamma_\text{ref}$, and set $6\pi\eta=1$, where $\eta$ is the dynamic viscosity of the suspending fluid.
The external shear rate, $\dot\gamma$, is varied from 0 to 3.
The rest of the simulation parameters are summarized in Tables \ref{tab:shakers}, \ref{tab:ABPs}, and \ref{tab:Brownian} for suspensions of shakers, active Brownian particles (ABPs) and passive colloids \xx{(not included in the main text)}, respectively.

Apart from the HIs and Brownian motion (if any), nearly touching particles are also subjected to a pairwise repulsive force, c.f.~Eq.~\eqref{eq:F_ext}.
With a reference force $F_\text{ref} \equiv 6\pi\eta a^2 \dot\gamma_\text{ref} =1$, we set $F_0/F_\text{ref} = 10$ and $\kappa^{-1} = 0.01$ across all simulations to model a strong, short-range repulsive force.
The repulsive force can also be used to define a time scale, $\tau_0 \equiv 6\pi\eta a^2/F_0 = 0.1$, which is the fastest physical time scale in our simulations.
We have ensured that $\tau_0$ is always at least one order of magnitude smaller than the other time scales.
For example, the time scale due to shaker activity, $\tau_a \equiv a/|B_2|$, is at least 50 times larger than $\tau_0$, and the time scale due to ABP self-propulsion, $a/B_1$, is at least 100 times larger.
The timestep for the temporal integration is fixed at $10^{-3}$ in all simulations.
The relative tolerance of the iterative solver for Eq.~\eqref{eq:final} is $10^{-4}$ in the active, noncolloidal cases and $10^{-3}$ in the passive, colloidal cases.
The choice of these numerical parameters is consistent with previous simulations \citep{fiore2019fast, Ge_Elfring_2025, Ge2022}.

\begin{table}
  \centering
  \setlength{\tabcolsep}{14pt} \renewcommand{\arraystretch}{1.2} 
  \begin{tabular}{cccccccc}
    \hline
    $\phi$   &   $k_B T$   &   $\tau_B$  \\
    \hline
    10\% -- 40\% & 0.001 & 1000\\  
    10\% -- 40\% & 0.01   & 100\\  
    10\% -- 40\% & 0.1     & 10\\  
    \hline
  \end{tabular}
  \caption{Parameters of the passive colloidal simulations, where $k_B T$ is the thermal energy, and $\tau_B \equiv 6\pi \eta a^3/(k_B T)$ the time scale due to Brownian motion as $\phi \to 0$.}
  \label{tab:Brownian}
\end{table}

\subsection{Solver verification}

Our numerical solver has been verified extensively for passive suspensions under shear \citep{Ge2022} and active suspensions without shear \citep{Ge_Elfring_2025}.
In the following, we provide two additional cases to verify Brownian motion.

\begin{figure*}
  \centering
  \includegraphics[height=7.8cm]{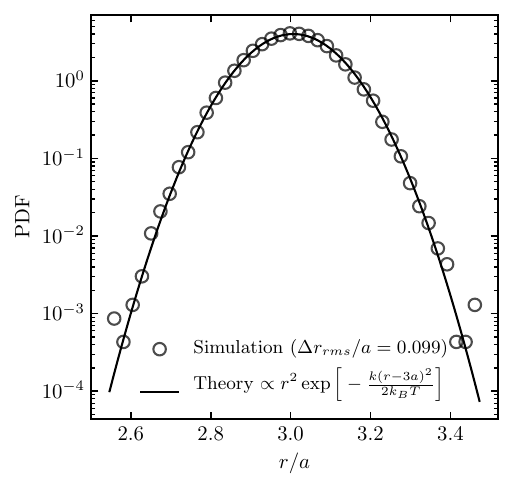}
  \includegraphics[height=7.8cm]{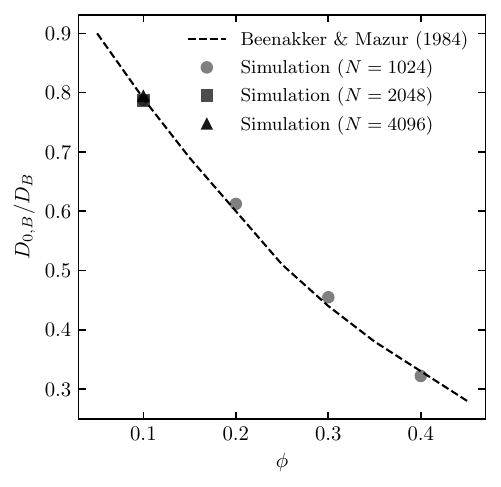}
  \caption{Verification of Brownian motion. 
  (Left) Distribution of the separation between two Brownian particles under an attractive potential.
  (Right) Short-time self-diffusion coefficient of Brownian suspensions at different volume fractions.}
  \label{fig:bro_pair}
\end{figure*}

First, we consider a pair of Brownian spheres under an attractive potential, $V(r)=\frac{k}{2}(r-r_{eq})^2$, where $k$ is the stiffness of the potential, and $r_{eq}$ the equilibrium separation.
Theoretically, the separation ($r$) between the two colloids of energy $k_\text{B}T$ each satisfies the Boltzmann's distribution, 
\begin{align}
    P(r) = \frac{r^2}{\sqrt{2\pi} \sigma (\sigma^2 + r_{eq}^2)} \exp \bigg[ -\frac{(r-r_{eq})^2}{2 \sigma^2} \bigg] ,
\end{align}
where $\sigma = \sqrt{k_\text{B}T/k}$.
We impose $k_\text{B}T=10^{-2}$, $k=1$, $r_{eq}=3$, and simulate in a cubic box of size $L=35$ with periodic boundary conditions.
The resulting probability density function (PDF), sampled over $t=50\tau_\text{B}$, where $\tau_\text{B} \equiv 6\pi\eta a^3/(k_\text{B}T)$, is shown in Fig.~\ref{fig:bro_pair} (Left).
We observe a close agreement between simulation and theory, verifying the Brownian calculation.

Next, we simulate passive colloidal suspensions at various volume fractions and examine the short-time self-diffusivity, \xx{$D_{0,B}$}.
For an isolated colloid, the self-diffusivity is given by the Stokes-Einstein relation, $D_B=k_\text{B}T/(6\pi\eta a)$.
In a suspension, \xx{$D_{0,B}/D_B < 1$} due to hydrodynamic and excluded-volume interactions.
Fig.~\ref{fig:bro_pair} (Right) shows our simulation results in comparison to the theory of \citet{Beenakker1984}, which considered many-body HIs.
Here, we have applied the correction proposed by \citet{ladd1990hydrodynamic} to account for the periodic effect.
The close agreement between simulation and theory, as well as the collapse of simulation results under different number of particles, verifies the implementation of Brownian motion under HIs.

\section{Results}
\label{sec:results}


\subsection{Shakers}

\begin{figure*}
  \centering
  \includegraphics[height=7.cm]{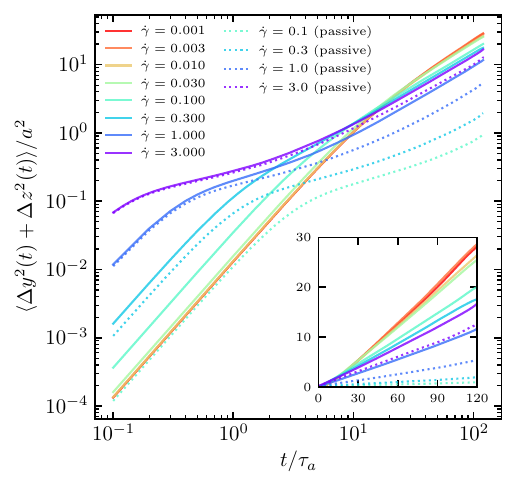}
  \includegraphics[height=7.cm]{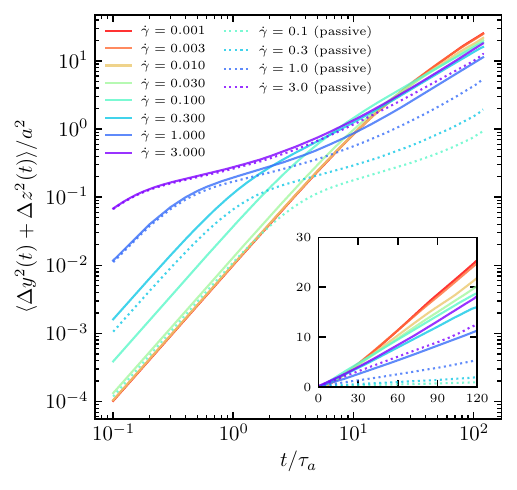} \\
  \includegraphics[height=7.cm]{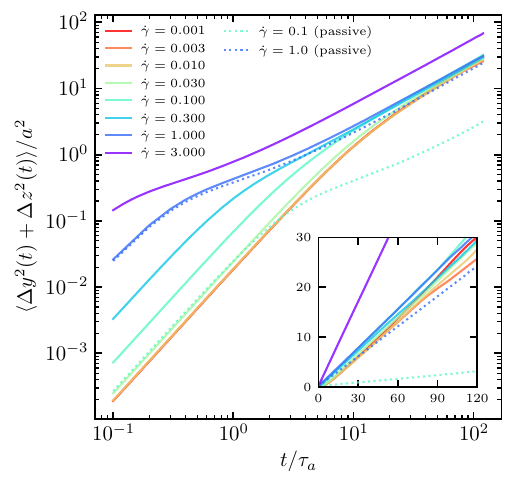}
  \includegraphics[height=7.cm]{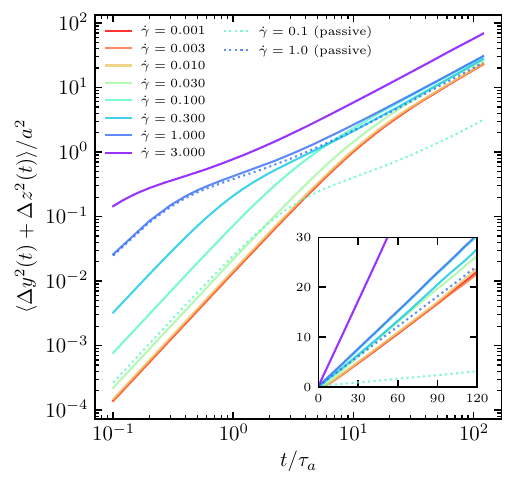} \\
  \includegraphics[height=7.cm]{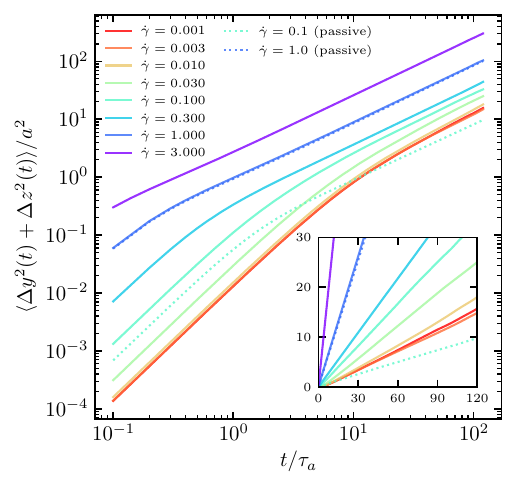}
  \includegraphics[height=7.cm]{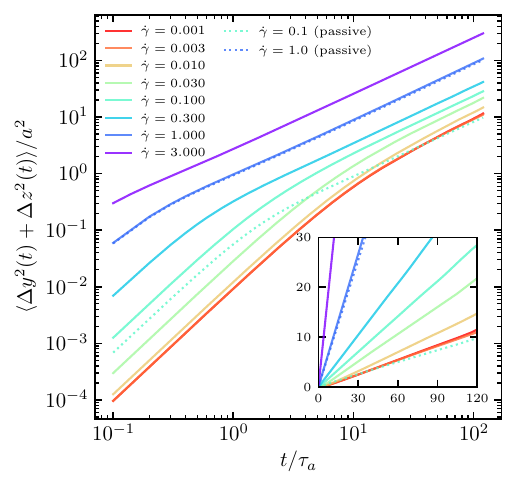} \\
  \caption{MSDs in the $yz$-plane for pulling (left column) and pushing (right column) shakers 
               under different shear rates ($\dot\gamma$, \xx{in raw simulation unit}) and volume fractions ($\phi$). 
               Top row: $\phi=10\%$; middle row: $\phi=20\%$; bottom row: $\phi=40\%$.
               Dotted lines are passive suspensions \xx{(without activity, or $B_2=0$)}.
               Insets show the same data on linear scales.}
  \label{fig:msd}
\end{figure*}

In the main article, we showed the mean square displacements (MSDs) for pulling shakers under different \xx{P\'eclet numbers (Pe)} at volume fraction $\phi=10\%$; c.f.~Fig.~1(b) therein. 
Fig.~\ref{fig:msd} here shows the MSDs at $\phi=10\%$, 20\% and 40\% for pullers, pushers and passive particles \xx{at different shear rates, $\dot\gamma$} (the results at $\phi=30\%$ are similar to those at 40\%).
We can make the following observations:
(i) at low $\dot\gamma$, the dynamics always displays a crossover from short-time ballistic to long-time diffusive motions, consistent with our previous simulations of shakers without shear \citep{Ge_Elfring_2025};
(ii) as $\dot\gamma$ increases, the MSD at short times tends to increase regardless of $\phi$; 
(iii) however, the MSD at long times can either decrease ($\phi=10\%$), remain relatively unchanged ($\phi=20\%$), or increase ($\phi=40\%$).
These different scaling behaviors are also illustrated in Fig.~1(c) in the main article ($D/D_0$ vs Pe).
We note that the transient subdiffusive motion, caused by shear and thus more pronounced at higher Pe, tends to \emph{diminish} as $\phi$ increases. 
This is consistent with the velocity anticorrelation shown in Fig.~2(b) in the main article, which is the reason for the varying scaling relationships between $D/D_0$ and Pe \xx{in the intermediate Pe range} at different $\phi$.

\subsection{ABPs}

\begin{figure*}
  \centering
  \includegraphics[height=7.8cm]{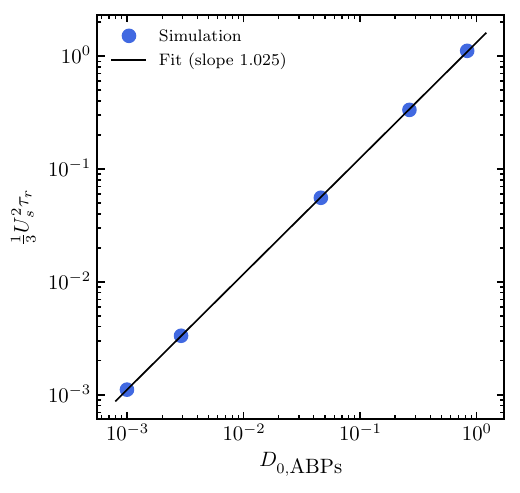}
  \includegraphics[height=7.8cm]{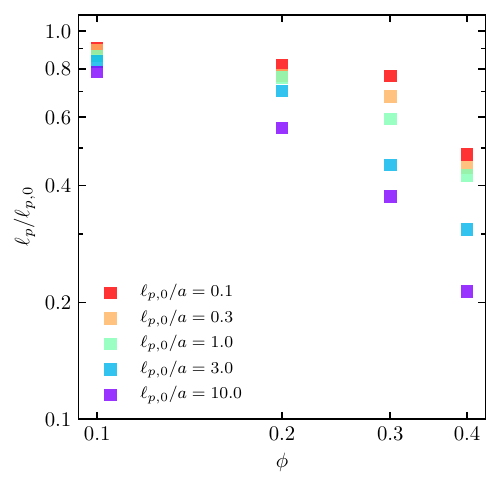}
  \caption{\xx{(Left)} Diffusivity of ABPs at different persistence lengths at $\phi=10\%$ without shear. Both $U_\text{s}$ and $\tau_r$ are input parameters. \xx{$D_{0,\text{ABPs}}$} is extracted from the MSDs.
               \xx{(Right) Ratio of the observed and single-particle persistence lengths, $\ell_p/\ell_{p,0}$, as a function of $\phi$.}}
  \label{fig:D0_ABPs}
\end{figure*}

\begin{figure*}
  \centering
  \includegraphics[height=7.5cm]{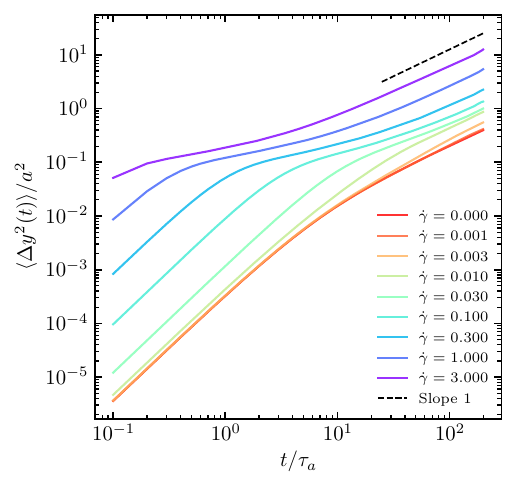}
  \includegraphics[height=7.5cm]{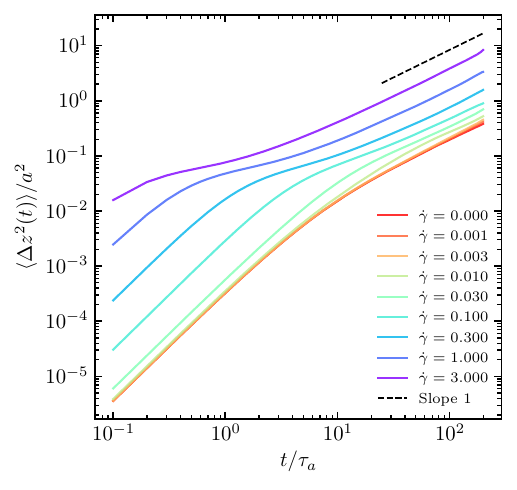} \\
  \includegraphics[height=7.5cm]{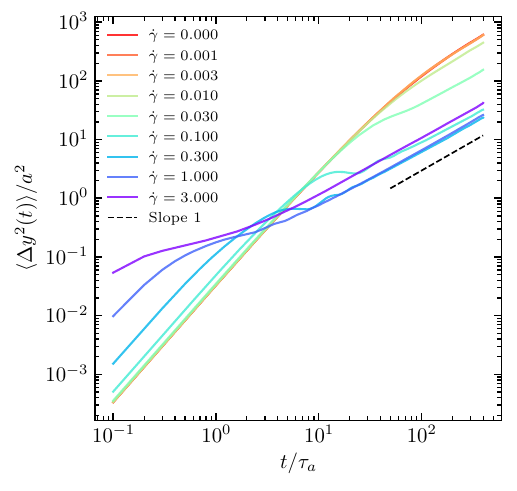}
  \includegraphics[height=7.5cm]{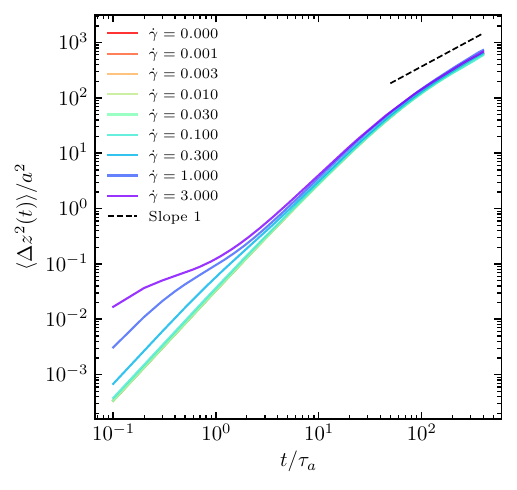} \\
  \caption{MSDs in the $y$ (left column) and $z$ (right column) directions of ABPs at $\phi=10\%$. 
               Top row: \xx{$\ell_{p,0}/a=0.1$}; bottom row: \xx{$\ell_{p,0}/a=10$}.}
  \label{fig:msd-ABPs}
\end{figure*}

In the main article, we mentioned that the underlying diffusivity of ABPs, \xx{$D_{0,\text{ABPs}} \to \frac{1}{3}U_s^2 \tau_r$ as $\phi \to 0$, where $U_s$ is the self-propulsion speed, and $\tau_r$ the rotational relaxation time.
The left panel of} 
Fig.~\ref{fig:D0_ABPs} verifies this relation at $\phi=10\%$.
\xx{Note that, both $U_s$ and $d_r$ are \emph{uniformly} imposed on all particles (because ABPs do not interact hydrodynamically due to their self-propulsion), while $D_{0,\text{ABPs}}$} is the \emph{observed} diffusion coefficient extracted from fitting the MSDs according to $\langle \Delta x^2(t) + \Delta y^2(t) + \Delta z^2(t) \rangle = 6 D_{0,\text{ABPs}} (t-t_0)$ in the long time limit ($t_0$ is generally nonzero as the dynamics is not diffusive at short times; see Fig.~\ref{fig:msd-ABPs}).
\xx{We also mentioned that the persistence length, $\ell_p \equiv 3D_{0,\textrm{ABPs}}/U_{rms}$, where $U_{rms}$ is the root-mean-square particle speeds, is capped at $\ell_{p,0} \equiv U_s\tau_r$ as $\phi \to 0$. 
The right panel of Fig.~\ref{fig:D0_ABPs} confirms that $\ell_p/\ell_{p,0} < 1$ is indeed satisfied for all simulated $\ell_{p,0}$ and $\phi$.
It also shows that $\ell_p$ tends to reduce further when $\ell_{p,0}$ is larger, due to the crowding effect.}

Fig.~\ref{fig:msd-ABPs} shows the separate MSDs in $y$ and $z$ directions at very small and large \xx{$\ell_{p,0} \equiv U_\text{s}\tau_r$ at $\phi=10\%$.
When $\ell_{p,0}/a \ll 1$,} both $\langle \Delta y^2(t) \rangle$ and $\langle \Delta z^2(t) \rangle$ increase with $\dot\gamma$, consistent with the expectation at smaller persistence.
Interestingly, when \xx{$\ell_{p,0}/a \gg 1$}, $\langle \Delta y^2(t) \rangle$ can not only reduce significantly with $\dot\gamma$, but also exhibit an oscillatory behavior.
This is a result of the external rotation mentioned in the main article.
As the rotational speed scales with the shear rate, $\Omega_\infty = \dot\gamma/2$, the period of the oscillation scales inversely with $\dot\gamma$, as can be verified.
On the other hand, $\langle \Delta z^2(t) \rangle$ remains persistent as the rotation is around the $z$ direction and ABPs hardly interact with each other unless they collide (which not only has a direct impact on the colliding pairs, but would also generate force dipoles that affect the dynamics of distant particles via HIs). 
In the case of \xx{$\ell_{p,0}/a=10$} as shown in Fig.~\ref{fig:msd-ABPs}, $\langle \Delta z^2(t) \rangle$ at long times is nearly independent of $\dot\gamma$ because \xx{$D_{0,\text{ABPs}}$} is large, thus the effect of self-propulsion dominates.

\begin{figure*}[t]
  \centering
  \includegraphics[height=7.8cm]{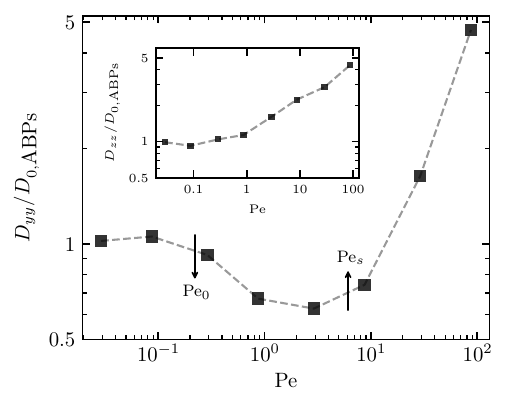}
  \caption{\xx{Diffusivity of ABPs at $\phi=40\%$ and $\ell_{p,0}/a = 10$ (the observed $\ell_p/a \approx 2.1$).}}
  \label{fig:ABP_Dyy40}
\end{figure*}

\xx{Finally, we mentioned in the main text that diffusion could be suppressed even in \emph{concentrated} suspensions, since $D_s(\dot\gamma_s,\phi) < D_{0,\textrm{ABPs}}(\tau_s,\tau_r,\phi)$ at sufficiently large $\ell_p$; c.f.~the inset in Fig.~4 therein.
Fig.~\ref{fig:ABP_Dyy40} verifies the above statement at $\phi=40\%$ and $\ell_{p,0}/a = 10$, where the observed $\ell_p/a \approx 2.1$; c.f.~\ref{fig:D0_ABPs} (Right).}

\subsection{Passive colloids}

\begin{figure}[t]
  \centering
  \includegraphics[height=8.5cm]{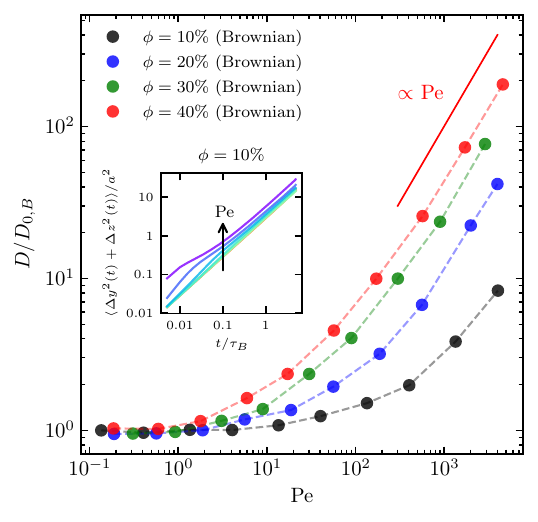}
  \caption{\xx{Diffusivity of passive colloidal suspensions at different $\phi$. 
  The inset shows the MSDs at $\phi=10\%$ under different Pe, 
  where $\tau_B \equiv 6\pi \eta a^3/\big(k_\text{B}T\big)$.}}
  \label{fig:BD}
\end{figure}

\xx{In the main text, we mentioned that the diffusivity of passive colloids ($k_B T > 0$) strictly increases with shear.
Fig.~\ref{fig:BD} shows the relative diffusivity, $D/D_{0,B}$, against Pe for $\phi$ from 10\% to 40\%, as well as the MSDs under different Pe at $\phi=10\%$.
Clearly, shear tends to enhance the diffusion at all $\phi$, despite the subdiffusive dynamics appearing transiently at higher Pe (see the inset).
Since passive colloids have zero persistence by definition (their fluctuations are uncorrelated), this suggests that a persistent internal motion is \emph{necessary} for observing a nonmonotonic diffusion under shear.}


\bibliographystyle{apsrev4-2}
\bibliography{supplement}